\documentclass[journal]{rmaa}

\title{Periodic Radio Continuum Emission Associated with the $\beta$ Cephei Star V2187 Cyg}

\author{Mauricio Tapia\altaffilmark{1}, 
  Luis F. Rodr\'\i guez\altaffilmark{2,4},
  Gagik Tovmassian\altaffilmark{1}, 
  Vicente Rodr\'\i guez-G\'omez\altaffilmark{3},
  Diego Gonz\'alez-Buitrago\altaffilmark{1}, 
  Sergei Zharikov\altaffilmark{1}, 
  and
  Gisela N. Ortiz-Le\'on\altaffilmark{2}}

  \altaffiltext{1}{Instituto de Astronom{\'\i}a, Universidad Nacional 
  Aut\'onoma de M\'exico, Campus Ensenada}
  \altaffiltext{2}{Centro de Radioastronom{\'\i}a y Astrof{\'\i}sica, Universidad Nacional 
  Aut\'onoma de M\'exico, Campus Morelia}
  \altaffiltext{3}{Harvard-Smithsonian Center for Astrophysics, USA}
  \altaffiltext{4}{Astronomy Department, Faculty of Science, King Abdulaziz University, Saudi Arabia}
  
\fulladdresses{
\item  Luis F. Rodr{\'\i}guez: Centro de Radiostronom{\'\i}a y Astrof{\'\i}sica, Universidad Nacional Aut\'onoma de 
M\'exico, A. P. 3-72, (Xangari), 58089 Morelia, Michoac\'an, M\'exico, and Astronomy Department, Faculty of 
Science, King Abdulaziz University, P.O. Box 80203, Jeddah 21589, Saudi Arabia 
(l.rodriguez@crya.unam.mx)
\item Gisela N. Ortiz-Le\'on: Centro de Radiostronom{\'\i}a y Astrof{\'\i}sica, Universidad Nacional Aut\'onoma de 
M\'exico, A. P. 3-72, (Xangari), 58089 Morelia, Michoac\'an, M\'exico, (g.ortiz@crya.unam.mx)
\item Mauricio Tapia, Gagik Tovmassian, Diego Gonz\'alez-Buitrago, Sergei Zharikov: 
Instituto de Astronom{\'\i}a, Universidad Nacional Aut\'onoma de 
M\'exico, Apartado Postal 877, Ensenada, Baja California, CP 22830, M\'exico  
(mt@astrosen.unam.mx, gag@astrosen.unam.mx, zhar@astrosen.unam.mx).
\item Vicente Rodr\'\i guez-G\'omez: Harvard-Smithsonian Center for Astrophysics, 
60 Garden Street, Cambridge, MA 02138, USA 
} 

\shortauthor{Tapia et al.}
\shorttitle{Periodic Radio Emission from V2187 Cyg}

\SetVolume{00} \SetFirstPage{001} \SetYear{2013}
\ReceivedDate{\today} 
\AcceptedDate{Year Month Day} 

\resumen{Presentamos fotometr{\'\i}a y espectroscop{\'\i}a de mediana resoluci\'on  de V2187 Cyg. Confirmamos que es 
una estrella tipo $\beta$ Cephei de tipo espectral B2-3~V. Tiene variaciones sinusoidales con un periodo de 0.2539 d{\'\i}as 
y amplitudes medias de 0.037 y 0.042 magnitudes en $i$ y $V$, respectivamente. No encontramos evidencia de variaciones 
en los perfiles de l{\'\i}neas de absorci\'on en escalas de tiempo de horas o d{\'\i}as y no muestra  l{\'\i}neas de emisi\'on. 
Reportamos tambi\'en la sorpresiva detecci\'on de emisi\'on de radiocontinuo de esta estrella con un espectro  t\'ermico. 
La emisi\'on de radio presenta variaciones con un periodo de 12.8 d{\'\i}as. De su archivo, encontramos que ninguna de las 
otras 15 estrellas de tipo  $\beta$ Cephei que han sido observadas con el Very Large Array muestra emisi\'on de radio
detectable, aunque muchas de ellas se encuentran mucho m\'as cercanas al Sol que V2187 Cyg. La
emisi\'on de radio es, entonces, un fen\'omeno extremadamente raro en las estrellas tipo  $\beta$ Cephei.}

\abstract{We present new optical time-resolved photometry and  medium-resolution spectroscopy of V2187 Cyg. We confirm its
classification as a $\beta$ Cephei star based on sinusoidal light variations with a period of 0.2539 days and mean amplitudes of 
0.037 and 0.042 magnitudes in $i$ and $V$, respectively.  We classified the spectrum of this star B2-3~V with no evidence of variations in 
the profiles of its absorption lines in timescales of hours or days. The stellar spectrum is totally absent of emission lines. We  detected 
unexpected faint radio continuum emission (between 0.4 and 0.8 mJy at 6-cm) showing a sinusoidal variation with a period of 12.8 days. 
The radio spectrum is thermal. We searched in the Very Large Array archive for radio continuum emission toward other 
15 $\beta$ Cephei stars. None of these additional stars, some of them much closer to the Sun than V2187 Cyg, was detected, 
indicating that radio emission is extremely uncommon toward $\beta$ Cephei stars.}
      

\begin{document}

\maketitle

\section{Introduction}

During analysis of the NRAO4\footnote{The National Radio Astronomy Observatory is a facility
of the National Science Foundation operated under cooperative agreement by Associated Universities, Inc.} 
Very Large Array (VLA) archive data of the O-type binary system
Cyg OB2 \#8A, the persistent presence of a faint radio source $\sim 1.3\arcmin$  to its southeast was noted. The 
position of this faint radio source is coincident, within $0.2\arcsec$, with the $\beta$ Cephei star
V2187 Cyg, (ICRS 2MASS coordinates  $20^{\rm h} 33^{\rm m} 18\rlap.{^{\rm s}}29; +41^o 17' 39\rlap.{''}4$),
in the central part of the spherical OB association Cyg~OB2 (Kn\"odlseder 2000).
The  $\beta$ Cephei stars are pulsating main sequence variables of low amplitude ($\Delta V \le 0.2$) with early 
B spectral types and typical periods of several hours (Stankov \& Handler 2005).

V2187 Cyg, also designated MT91-487 (Massey \& Thompson 1991) or Schulte-63 (Schulte 1956), was classified
by Pigulski \& Kolaczkowski (1998) as a member of the $\beta$ Cephei class of variables based on its 
observed (Kron-Cousins) $I$-band sinusoidal light curve with an amplitude of 0.033 magnitudes, a derived period of 0.25388 
days and its photometric color indices ($V=15.39, U-B=0.66, B-V=1.61, V-R=1.14$ and $V-I=2.14$; Massey \& Thompson 1991;
Pigulski \& Kolaczkowski 1998) that indicate an early-B spectral type. Its 2MASS near-infrared magnitudes are 
$J=11.52, H=10.97, K=10.67$. Albacete-Colombo et al. (2007) list this star as an X-ray emitter.

\begin{figure*}[htb]
\centering
  \includegraphics[width=1.0\columnwidth]{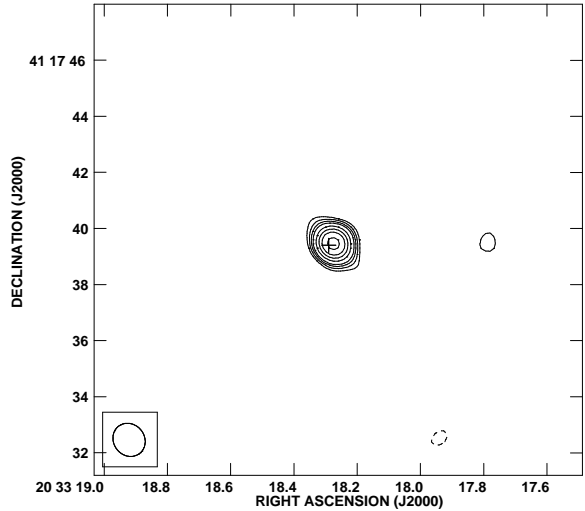}
 \caption{VLA image of the radio emission associated
 with V2187 Cyg at 6-cm taken on 2005 Feb 13. Contours are -4, -3,  3, 4, 5, 6, 8, 10, 12 and 15
times 45 $\mu$Jy, the rms noise of the image.
The synthesized beam, shown in the bottom left corner,
 has half power full width dimensions of
 $1\rlap.{''}23 \times 1\rlap.{''}10$,
 with the major axis at a position angle of $+38^\circ$.
 The cross marks the optical position of the star.}
   \label{fig1}
   \end{figure*}

\begin{figure*}[htb]
\centering
  \includegraphics[width=1.85\columnwidth]{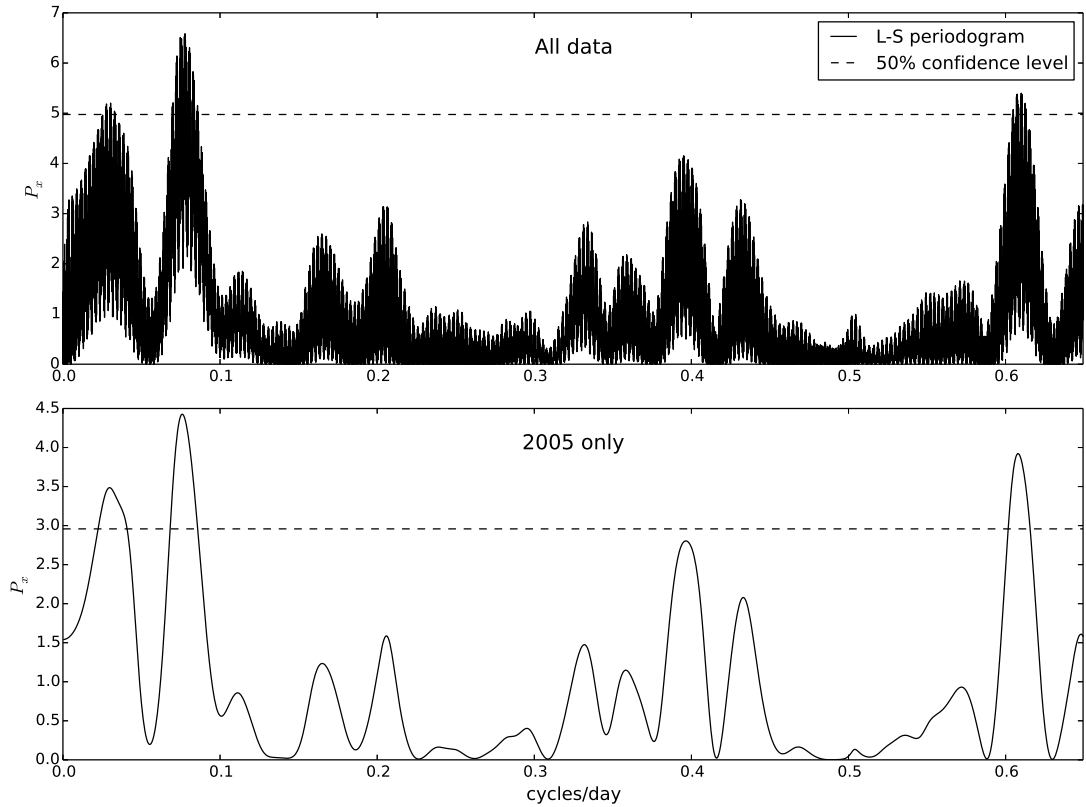}
  \caption{Periodograms for the radio continuum data of V2187 Cyg. The upper panel is for
  all 17 data points; the high frequency oscillation is produced by the inclusion of data points 
  remote in time from the 2005 observations. The lower panel is for the 13 points obtained
  during a 36-day span in 2005 February-March. The dashed lines indicate the 50\% confidence level.}
\label{fig2}
\end{figure*}

\begin{figure*}[htb]
\centering
  \includegraphics[width=1.3\columnwidth]{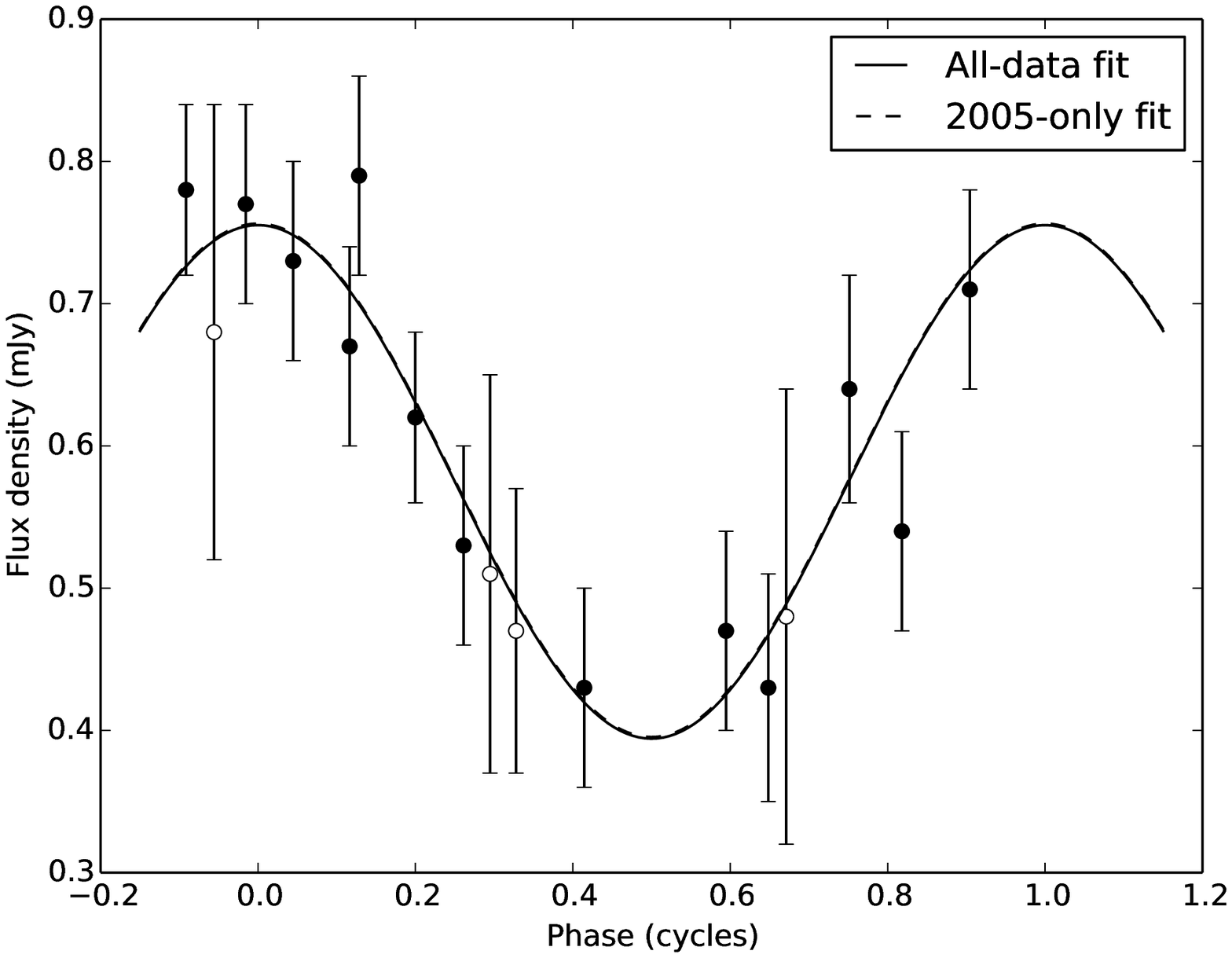}
  \caption{VLA observations of V2187 Cyg at 6-cm. The flux densities are
  plotted as a function of orbital phase in the 12.78-day period,
  with phase 0 corresponding to the maximum radio emission.
  The data points from 2005 are shown as solid circles, while the data points for
  other epochs are shown as empty circles. The continuum line shows the cosine fit for
  all data points and the almost identical broken line represents the fit for the 13
  data points obtained in 2005.}
\label{fig3}
\end{figure*}

The star is located close to the highly obscured ($A_V = 5 -10$) central region of the Cyg OB2 association that is characterized 
by an extremely high projected stellar density, particularly rich in O and early-type stars. The whole association, or young galactic
globular cluster, contains around 2600  O and B stars, with a total mass amounting to slightly less than $10^5 M_\odot$ 
(Kn\"odlseder 2000). The age of the complex  has been estimated by several authors  to be in the range of 3 - 4 million years old
(Torres-Dodgen et al. 1991, Herrero et al. 1999, Kn\"odlseder  et al. 2002).  Cyg OB2 is also conspicuous in terms of harboring a large
fraction of known massive (OB- type) binaries. Kimimki et al. (2012) report orbital and stellar parameters of 20 of these
spectroscopic binary systems.  More intriguing is the presence of three known X-ray sources and (variable) non-thermal stellar radio emitters 
close to the nucleus of Cyg OB2. They are Cyg OB2 \#9,  \#8a  and \#5, all of these being O-type 
spectroscopic binaries (Naz\'e et al. 2012, De Becker et al. 2004), the latter  with at least two other less massive stellar companions 
(Dzib et al. 2013 and references therein). 

There is no known report of radio continuum emission from stars classified as $\beta$ Cephei
(G\"udel 2002; Wendker 2004).  Their lack of radio continuum emission is consistent with what is known regarding early type
stars, that makes O-type stars much stronger radio emitters than B-type stars. Thermal radio continuum
in massive stars is due to free-free emission in their ionized winds, that is proportional to the mass loss
rate to the (4/3) power (e.g. Blomme 2011). O-type stars have much more massive winds than
B-type stars and are, thus, easier to detect in the radio continuum (e.g. Bieging et al. 1989).
Non-thermal radio continuum emission can also be present in early type stars and it is known to be associated
with binarity (Dougherty \& Williams 2000). In a massive stellar binary, the two stellar winds
collide and in the shock a fraction of the electrons is Fermi accelerated to relativistic speeds, producing
synchrotron emission (De Becker 2007; Blomme 2011). The strength of the radio emission increases with the 
mass loss rate and wind speed, and again O-type stellar binaries are frequently detected as radio
emitters. All the non-thermal radio stellar sources listed by De Becker (2007) contain at least one O-type
or Wolf-Rayet (WR) star.

In this paper we report new optical photometry in the $i$ and $V$ bands and time-resolved medium-resolution optical 
spectroscopy of V2187 Cyg. From these, we classify the star as B2-3~V and derive its basic stellar properties, that are
consistent with it being a peculiar $\beta$ Cephei star. To understand the nature of the radio emission from V2187 Cyg, we 
analyze flux densities at 6-cm from the VLA  archive data  at 17 epochs from 1984 to 2005, obtaining evidence of
periodic variations in the radio-continuum emission. We investigate the nature of this  emission 
and also perform a search in the VLA archives  for radio emission from other  $\beta$ Cephei stars.

\section{Optical Observations}

Medium-resolution long-slit spectroscopy was performed with the Boller \& Chivens spectrograph 
(http://www.astrossp.unam.mx) attached to the 2.1-m telescope of the Observatorio Astron\'omico
Nacional at San Pedro M\'artir (OAN-SPM) of Mexico during three runs in 2011 June, 2012 July and 2012 September. 
We used the 1200 l/mm grating and a slit width of $250~\mu$m, yielding a nominal spectral resolution of 1.8 \AA~ 
(111 km~s$^{-1}$ at H$\beta$).  This spectral resolution was confirmed by measuring the width 
of the interstellar Na D lines on all the spectra of stars. The scale on the Marconi \#2 CCD chip that was 
used is 0.60  \AA /pixel. The length of the slit is 55\arcsec~ oriented east-west. Two stars were included 
in the slit:  V2187 Cyg and the O9V star MT91-507 (Massey and Thompson 1991), 
located 30\arcsec~ to the east, and which we used as comparison. 
Series of spectra in the blue (4000 - 4900 \AA) and red (5570 - 6800 \AA) wavelength regions
were secured. Table 1 presents the details of the individual spectroscopic 
observations. Note that the integration times were kept relatively short in order to search 
for possible short-time (fractions of an hour) variations. All spectra were reduced in the 
standard way (except no flat-fielding was performed) using IRAF's procedures\footnote{IRAF is distributed by
the National Optical Astronomy Observatory, which is operated by the Association of Universities for Research in Astronomy, Inc. under
contract to the National Science Foundation.}. Wavelength calibrations were performed 
using frames of CuArNe arcs taken several times each night. 

\begin{figure*}[htb]
\centering
  \includegraphics[width=1.5\columnwidth]{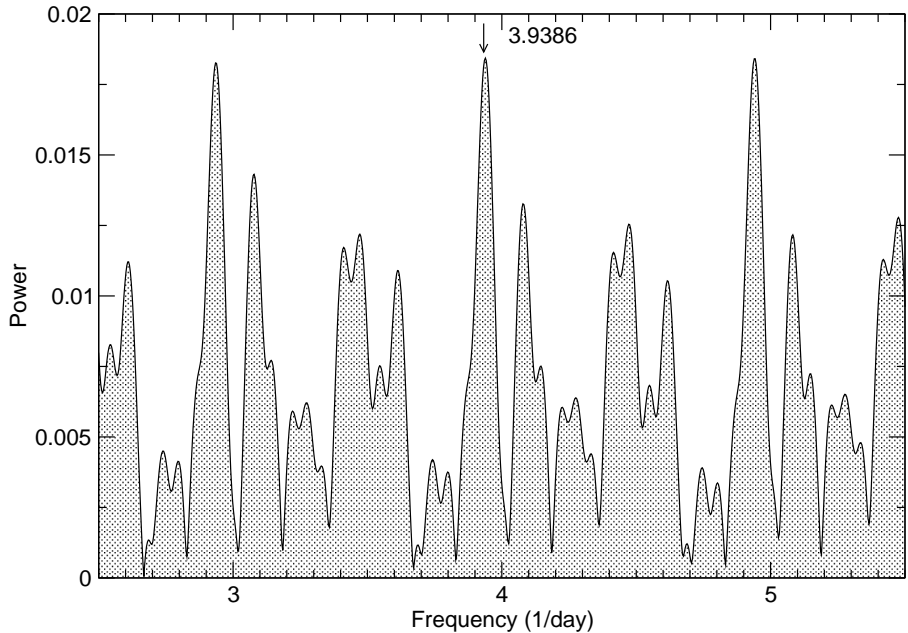}
  \caption{The power spectrum of the $i$-band variation of V2187 Cyg.}
\label{fig4}
\end{figure*}

\begin{figure*}[htb]
\centering
  \includegraphics[width=1.4\columnwidth]{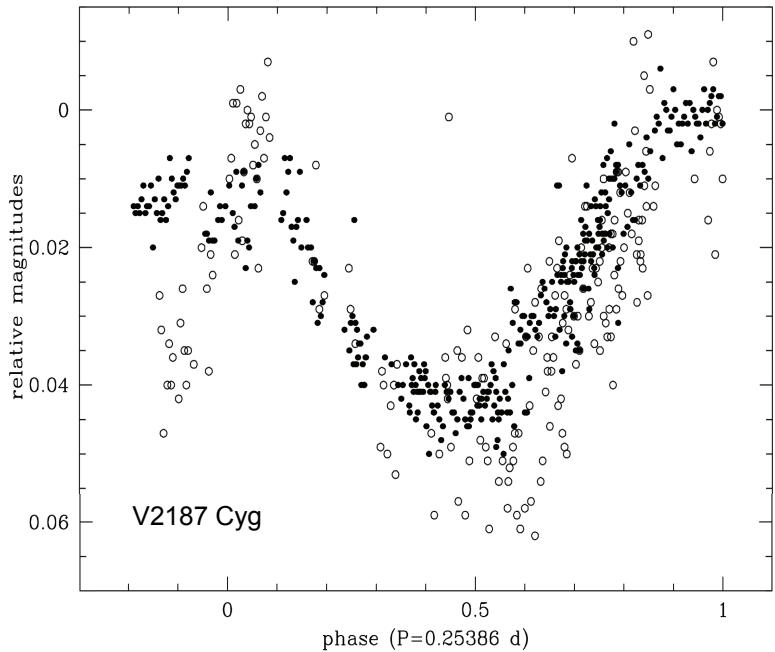}
 \caption{Light curves of V2187 Cyg in $i$ (filled circles) and $V$ (open circles). The scale in the abscissa is given in terms of arbitrary 
phase with a period = 0.2539 days (see text). Measurements from the first  observing night in this series correspond to the 
right hand side of the plot, while those from the last night, to the left (see text). The scale in the ordinate is expressed in relative magnitudes. 
Note that the significantly larger amplitude and scatter (noise) in the $V$-band and the larger number of points for phases larger than 0.3, 
give  the false illusion of a phase shift.}
   \label{5}
   \end{figure*}

\begin{figure*}[htb]
\centering
  \includegraphics[width=1.5\columnwidth]{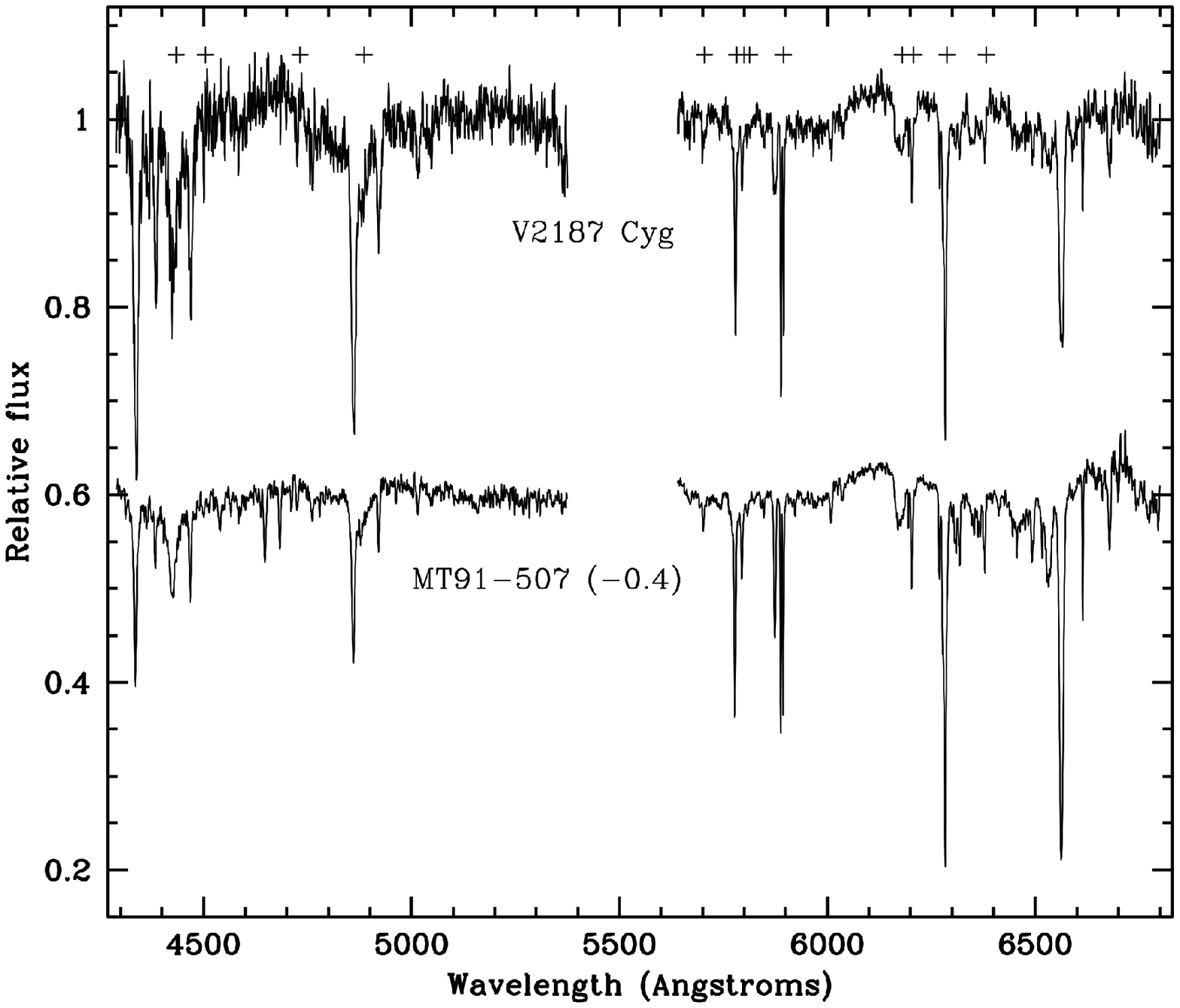}
 \caption{Averaged normalized spectra of V2187 Cyg and MT91-507.  The diffuse interstellar bands (DIBs) are indicated with 
plus signs at the top of the graph. Note that H$\beta$ blends with DIB4882. The dates of each averaged short wavelength segment are
2012Sep13+2011June29-31 and for the long wavelength segment, 2012Sep14-16+2012July25-27+2012July18-2.}
   \label{f6}
   \end{figure*}

We also performed a series of  photometric measurements of V2187 Cyg and several nearby comparison stars within 
50\arcsec ~on 60-second integration images. These were acquired in  approximately 1.5-hour nightly periods from 
2012 October 23 to 2012 November 7, totalling nearly 14 hours of coverage. The frames were obtained simultaneously 
through the  Bessel $V$ and $i$ filters. The instrument used was a 4-channel optical/near-infrared camera (though only the two 
optical cameras were used for this work) named RATIR (Reionization and Transients Infrared Camera) 
attached to the 1.5-m telescope of OAN-SPM (Butler et al. 2012; see also http://www.ratir.org). 
All observations were made in service mode. The typical measured full-width half maximum (FWHM) seeing was 
$1\rlap.{''}5$. No attempts were made to obtain absolute photometry and 
only differential photometry is reported in this paper. This is referred to several local comparison, non-variable 
(Pigulski \& Kolaczkowski 1998) stars. Digital aperture photometry was performed using DAOPHOT (Stetson1987) with an 
aperture (diameter) of 3\arcsec, large enough to allow for ocasional guiding errors.  Sky subtraction was performed  
in a ring of radius $1\rlap.{''}6$ and width $1\rlap.{''}1$. For some 
5-10\% of the usable time, thin clouds were present. The  photometric uncertainties were estimated from the 
dispersion of the relative magnitude differences between comparison stars. These values ranged from 
0.005 to 0.008 mag for $i$ and approximately three times larger for $V$ band.  The mean $V$ magnitude differences between 
the comparison O9V star MT91-507 and V2187 Cyg is 2.65, in full agreement with Massey and Thompson's 
(1991) photometry for these stars.

The photometric observations cover all phases of the short-period (0.25388 days = 6.1 hours)
sinusoidal variations reported for V2187 Cyg by Pigulski \& Kolaczkowski (1998). The star was observed 
for up to 1.6 hour-long segments each night during the length of our observing run (16 consecutive nights) and 
we manage to cover all phases, but this situation made it impossible to detect unequivocally any other photometric 
fluctuation in time-scales of several days.

\begin{figure*}[htb]
\centering
  \includegraphics[width=1.6\columnwidth]{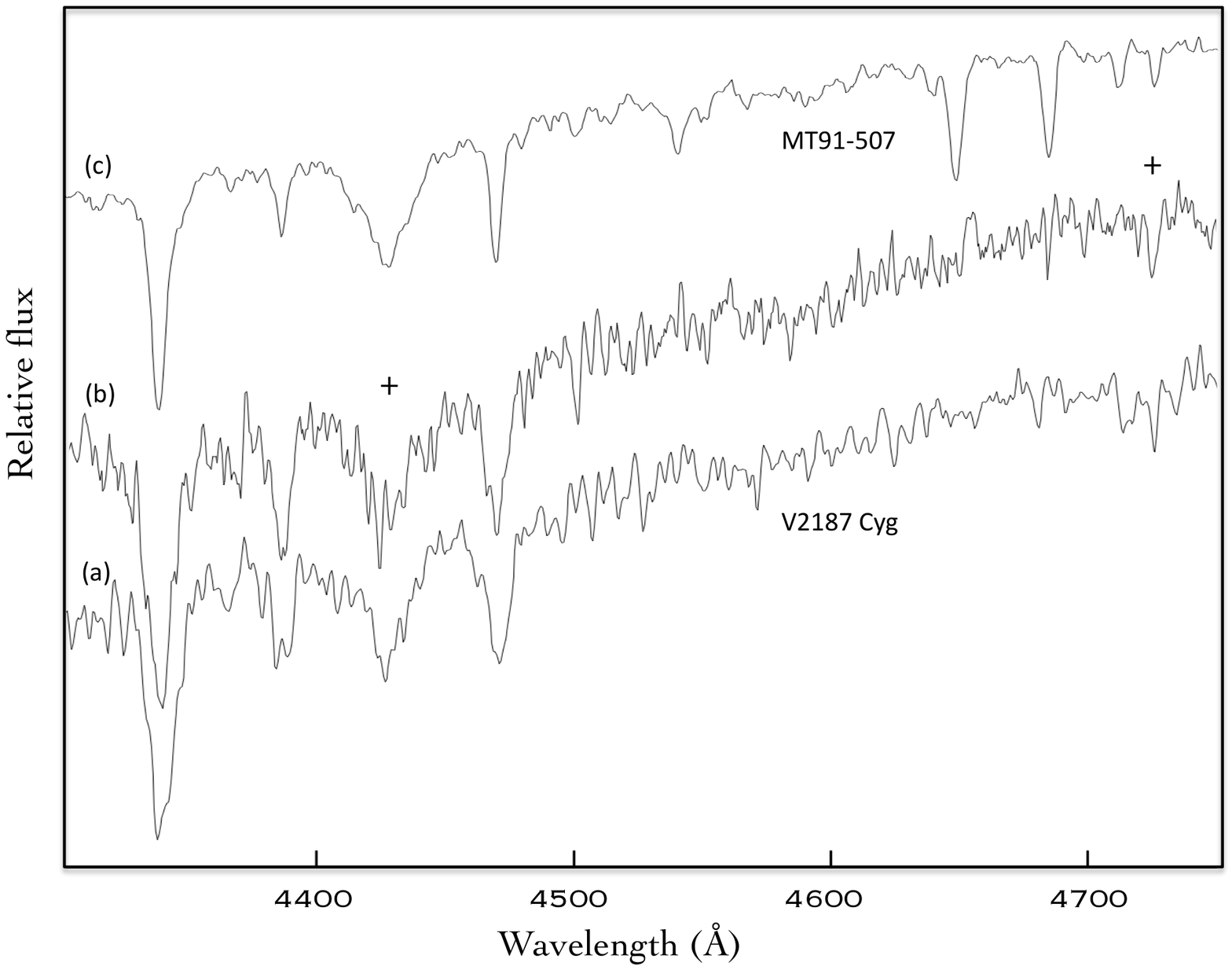}
 \caption{Sections of 3-day averaged spectra (boxcar-smoothed with kernel  3) of V2187 Cyg and MT91-507 with representative He and H lines at different epochs
and phases $\phi$ (Table~4). The absorption lines in the 4300-4750\AA~ segment are: H$\gamma$, HeI 4385,4471,4538,4713 CIII 4647/ 50/52,  HeII 4686, DIB4424,4724,4761. 
The dates are (a) 2011 June 29-31  ($\phi=0.81$); (b) 2012 September 13  ($\phi=0.48$);   (c) 2012 September 13. The plus signs mark the DIBs.}
   \label{f6}
   \end{figure*}

\begin{figure*}[htb]
\centering
  \includegraphics[width=1.6\columnwidth]{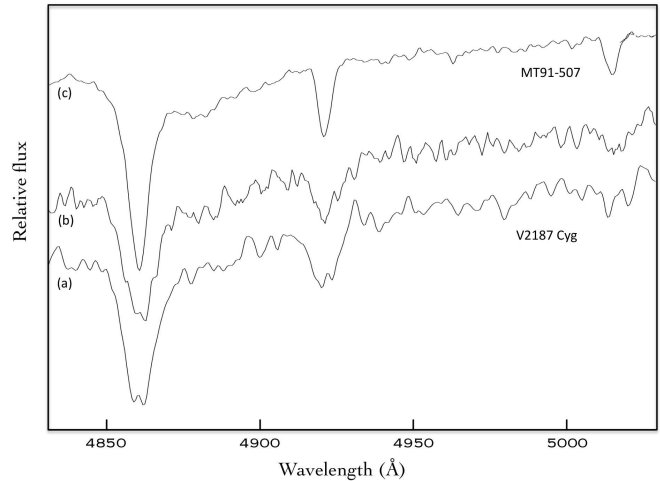}
 \caption{Sections of 3-day averaged spectra (boxcar-smoothed with kernel  3) of V2187 Cyg and MT91-507 with representative He and H lines at different epochs
 and phases $\phi$ (Table~4). The absorption lines in the 4830-5040 \AA~ segment are: H$\beta$, HeI 4920,5016.  The dates are  (a) 2011 June 29-31  ($\phi=0.81$); 
(b) 2012 September 13  ($\phi=0.48$);   (c) 2012 September 13.}
   \label{f7}
   \end{figure*}

\begin{figure*}[htb]
\centering
  \includegraphics[width=1.6\columnwidth]{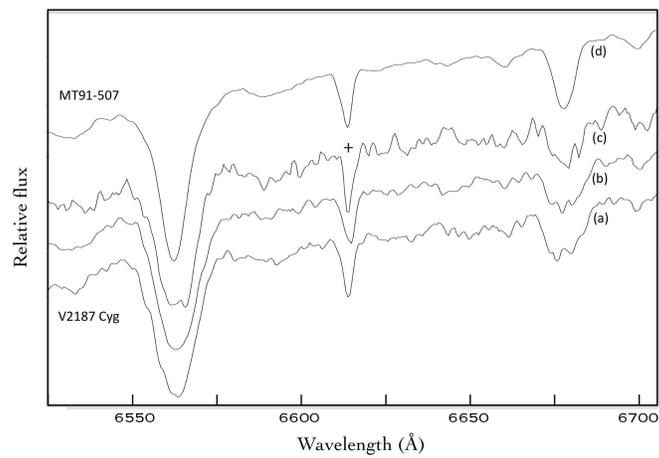}
 \caption{Sections of 3-day averaged spectra (boxcar-smoothed with kernel  3) of V2187 Cyg and MT91-507 with representative He and H lines at different epochs 
 and phases $\phi$ (Table~4). The absorption lines in the 6540-6700 \AA~ segment are: H$\alpha$, HeI 6672, DIB 6614. The dates are (a) 2012 July 18-21  ($\phi=0.17$); 
(b) 2012 July 25-27  ($\phi=0.32$); (c) 2102 September 14-16  ($\phi=0.63$); (d) All 2011-2012.  The plus sign marks the DIB.}
   \label{f8}
   \end{figure*}

\section{Results}

\subsection{6-cm Flux Density as a Function of Time} 

\begin{table*}[htbp]
\small
  \tablecols{4} 
\caption{\label{tab1} Log of the spectroscopic observations with the B\&Ch spectrograph on 2.1m telescope at SPM}
  \begin{center}
      \begin{tabular}{cccc}\hline\hline
      Date & No. spectra &  spectral range  &  exposure time (s)   \\
      UT & &  (\AA) & (s) \\ 
\hline
2011 June 29-31 & 4  &    4200 - 5300   &   900 \\
2012 July 18-27 & 38  &    5600 - 6800   &   900 \\
2012 Sept. 13 & 3       &   4300 - 5400   &   1200 \\
2012 Sept. 14-16 & 12 &  5600 - 6800   &   1200 \\
\hline\hline
  \label{tab1:1}
   \end{tabular}
\end{center}
\end{table*}

The bulk of the archive observations of V2187 Cyg comes from the
high-quality set of 13 observations used by Blomme et al. (2010)
to study Cyg OB2 \#8A. These observations were made at 6-cm 
between 2005 February 15 and 2005 March 12. Fig.~1 presents a representative VLA map 
from which measurements were performed. The flux densities from these observations
suggest periodic variations in the flux density. A chi-square test on these 13 data points 
yields a 94\% confidence level for such variability. We found four additional observations 
(two made in 1984, one in 1993 and one in 2004) of good quality made at 6-cm.
These observations are also included in Table 2.

\begin{table*}[htbp]
\small
  \setlength{\tabnotewidth}{0.9\textwidth} 
  \tablecols{7} 
\caption{\label{tab2} Flux densities of V2187 Cyg at 6-cm}
  \begin{center}
      \begin{tabular}{ccccccc}\hline\hline
      & JD & & Phase & Phase Calibrator & VLA & Source    \\
    Epoch & -2440000$^a$ &  Project & Calibrator & Flux Density (Jy) & Conf.  & Flux Density (mJy)$^b$ \\ 
\hline
1984 Mar 04 & 05764.36 & AA28   & 2007+404 & 4.23$\pm$0.01 & CnB & 0.51$\pm$0.14 \\
1984 Mar 09 & 05769.17 & AA28   & 2007+404 & 4.31$\pm$0.01 & CnB & 0.48$\pm$0.16 \\
1993 May 01 & 09109.11 & AS483  & 2007+404 & 3.19$\pm$0.01 & B   & 0.68$\pm$0.16 \\
2004 Feb 15 & 13051.28 & AS786  & 2015+371 & 2.76$\pm$0.01 & CnB & 0.47$\pm$0.10 \\
2005 Feb 05 & 13406.51 & AB1156 & 2007+404 & 2.37$\pm$0.01 & BnA & 0.67$\pm$0.07 \\ 
2005 Feb 06 & 13408.36 & AB1156 & 2007+404 & 2.44$\pm$0.01 & BnA & 0.53$\pm$0.07 \\ 
2005 Feb 08 & 13410.32 & AB1156 & 2007+404 & 2.46$\pm$0.01 & BnA & 0.43$\pm$0.07   \\
2005 Feb 11 & 13413.31 & AB1156 & 2007+404 & 2.48$\pm$0.01 & BnA & 0.43$\pm$0.08   \\
2005 Feb 13 & 13415.48 & AB1156 & 2007+404 & 2.36$\pm$0.01 & BnA & 0.54$\pm$0.07   \\
2005 Feb 18 & 13420.36 & AB1156 & 2007+404 & 2.37$\pm$0.01 & B & 0.62$\pm$0.06   \\
2005 Feb 25 & 13427.41 & AB1156 & 2007+404 & 2.43$\pm$0.01 & B & 0.64$\pm$0.08   \\
2005 Feb 27 & 13429.42 & AB1156 & 2007+404 & 2.42$\pm$0.01 & B & 0.78$\pm$0.06   \\
2005 Feb 28 & 13430.39 & AB1156 & 2007+404 & 2.40$\pm$0.01 & B & 0.77$\pm$0.07   \\
2005 Mar 01 & 13431.16 & AB1156 & 2007+404 & 2.43$\pm$0.01 & B & 0.73$\pm$0.07   \\
2005 Mar 02 & 13432.23 & AB1156 & 2007+404 & 2.41$\pm$0.01 & B & 0.79$\pm$0.07   \\
2005 Mar 08 & 13438.19 & AB1156 & 2007+404 & 2.41$\pm$0.01 & B & 0.47$\pm$0.07   \\
2005 Mar 12 & 13442.15 & AB1156 & 2007+404 & 2.44$\pm$0.01 & B & 0.71$\pm$0.07   \\
     \hline\hline
  \tabnotetext{a}{At the middle of the observing run.}
    \tabnotetext{b}{Corrected for the primary beam response.}
  \label{tab2:1}
   \end{tabular}
\end{center}
\end{table*}

We used the Lomb-Scargle method (Scargle 1982; Lomb 1976; Press et al. 1992)
to search for periodicity in the radio continuum emission (at 6-cm) in two ways:  
1) Using all 17 data points available, and 2) using only the 13 points obtained in a 36-day 
span in February-March 2005. Both periodograms are shown in  Fig.~2 and  suggest 
the same period, namely $12.8 \pm 0.2$ days. The false alarm probabilities, calculated assuming 
that the errors in the flux are Gaussian-distributed (Press et al. 1992), are 12.9\% (17 points) 
and 14.5\% (13 points). The cosine function derived from both fits are shown in Fig. 3. 
They are practically identical and have the same semi-amplitude of 0.18 mJy.  
By including measurements from earlier epochs we confirmed that the periodogram is 
dominated by the data from 2005. However, we note that 
the two measurements taken in 1984 have a clear significance to our model, 
since they are only separated by a few days. While these two measurements alone are not enough to 
firmly establish the existence of a periodicity, they certainly do not contradict our model, as one could 
think of possible configurations in which the two points cannot be fit by a 12.8-day sinusoid of the same amplitude 
(for example, if the two measurements had a flux approximately equal to the maximum value of our model and were 
temporally separated by 0.5 cycles, or if they had an average flux and were separated by 0.25 or 0.75 cycles).

\subsection{Radio Spectral Index}

Simultaneous observations are required to determine the spectral index. 
We found observations of V2187 Cyg in the VLA archive at 6~cm (4.86~GHz) and 3.6~cm (8.44~GHz) made 
on two epochs: 1993 May 01 as part of project AS483 and 2004 February 15 as part of project AS786. 
The flux densities at 4.86 GHz are given in Table 2, while those  
at 8.44 GHz are given in Table 3. Combining the flux densities at
both frequencies measured on 1993 May 1, we obtain an spectral index of $S_\nu \propto \nu^{0.8\pm0.5}$,
while for 2004 February 15, we obtain $S_\nu \propto \nu^{1.6\pm0.4}$. Both determinations 
are consistent with a thermal emission mechanism, most likely partially thick
free-free radiation. The other two flux densities listed
in Table 3 (at 20 and 12 cm) are also consistent with a thermal spectrum (i. e. weaker flux densities at
the lower frequencies).

\subsection{Optical photometry}

The Discrete Fourier Transform (DFT) method\footnote{http://www.univie.ac.at/tops/Period04/} was used  to
independently confirm the present periodicity of V2187 Cyg in the $i$-band photometric light curve. The resulting 
power  spectrum is shown in Fig.~4. The dominant frequency in this power spectrum is $3.93866\pm0.03503$
day$^{-1}$, which corresponds to a period of $0.25389\pm0.00228$. This value  agrees with that 
(0.25388 days) determined by Pigulski \& Kolaczkowski (1998), obtained in mid-1996 and based on a longer 
time-span (44 days) light-curve.

The light-curves of V2187 Cyg in $i$ (filled circles) and $V$ (open circles) are shown in Fig.~5. The photometric data points are
plotted folded into phase, assuming a period of 0.2539 days, covering 1.2 periods. The amplitudes of the light curves 
appear to vary, with values ranging from 0.032 to 0.042 magnitudes in $i$ and 0.035 to 0.050 in $V$ from one period to 
the next. In particular, one of the light maxima (phases 0.85 to 1.0 on the left hand extreme of the plot) on 2012 November 5-7
appear to be depressed compared to the equivalent phases on 2012 October 23 - 24 (on the right hand extreme of the plot). 
Based on these  light curves, and our determination of a B3~V spectral type, (see next subsection)  we  support
the classification of V2187 Cyg  as a member of the $\beta$~Cephei class of variables.

\subsection {Optical spectroscopy} 

We determine spectral types for V2187 Cyg and for the comparison star  MT91-507 (O8.5V, Hanson 2003; O9V, Kiminki et al. 2007)
using all available spectra. The spectrum of V2187 Cyg is classified for the first time and it yields B2-3~V,  based on the relative strength of the 
Balmer and HeI lines, using templates from Grey \& Corbally (2009), though a luminosity class IV cannot be ruled
out from our  spectrum. As a test of consistency, we used the same classification procedure to 
independently classify MT91-507, yielding O9V, in agreement with the literature. No evidence of emission lines in the spectrum 
of either V2187 Cyg or MT91-507 was found. The combined normalized spectra (all individual spectra added for each wavelength range) 
of both stars are shown in Fig.~6. 
The width (FWHM) of the absorption lines measured on this combined spectrum of V2187 Cyg  are $\sim 11-12$ \AA~ 
for the Balmer lines and  $\sim 8-10$ \AA~ for the HeI lines. The latter imply rotational velocities 
(v~sin~$i$) of around 350 km~s$^{-1}$ (Daflon et al. 2007, their Table 2). Those of MT91-507 are significantly narrower 
(around  9 \AA~and 5 \AA for H and HeI, respectively). 

Because of the low signal-to-noise (S/N), we found it impossible to obtain reliable values for the width of the absorption 
lines on individual short-integration (20~min)  spectra,  thus preventing us from determining whether there are variations
in the line profiles in time-scales comparable to the pulsation period (6 hours).
In order to search for evidence of line profile variations on timescales of days, we improved the S/N by averaging spectra 
of V2187~Cyg obtained over three consecutive nights in four epochs, which correspond to four distinct phases distributed along 
the 12.8 days radio period. These mean spectra, compared to those of the star MT91-507, are illustrated in Figs. 7, 8 and 9.
We then measured the widths of the Balmer and HeI lines in V2187~Cyg corresponding to each of the four epochs.
With typical uncertainties of $\pm 0.3$ for the HeI lines and $\pm 0.1$ for the H lines, the results are given in Table 4. 
Although there may be a marginal evidence of variations in phase with the radio-continuum flux, the large uncertainties 
preclude us from reaching any firm conclusion in this regard.

\section{Discussion}

We provide independent confirmation that this B3~V star displays the characteristics that define the $\beta$ Cephei class of pulsating 
variables, with a visual photometric period of 0.25388 days that has remained unchanged since its discovery in 1996 (Pigulski \& 
Kolaczkowski 1998). The measured widths of the HeI lines suggest atypically high rotation velocities, of about 300-400 km~s$^{-1}$. 
At the same time, the star shows a 12.8-day sinusoidal variation of its 6cm radio flux density, with a spectral index that 
implies a thermal origin. This kind of emission from V2187 Cyg is unique among $\beta$ Cephei stars, both in terms of its high radio 
flux as well as of its cyclic, variable nature, with a  period some 50 times longer than the pulsation period.

The available $UBVRIJHK$ photometry of V2187 Cyg combined with our new spectral type B2-3~V, imply a single value for the dust 
extinction of $A_V = 5.7-5.8$, consistent with that measured towards other {\it bona fide} members of the Cyg OB2 association. Determining
its spectroscopic distance from the Sun is subject to a large degree of uncertainty, mainly because of the uncertainties in the calibration of the 
stellar absolute magnitudes that must be assumed. It is probably more instructive to assume that V2187 Cyg  belongs to Cyg OB2
and is, thus, at the same distance. We adopt 1.62 kpc (true distance modulus 11.05) for the distance to Cyg OB2, as determined by 
VLBA astrometric parallax to two {\it bona fide} Cyg OB2 members (Zhang et al. 2012, Dzib et al. 2013). This value  is consistent with the mean 
spectroscopic  distance measurements ($d=1.7$ kpc; Torres-Dodgen et al. 1991, Massey \& Thompson 1991). The assumed distance implies absolute
magnitudes $M_V= -1.5$ and $M_K=-0.9$ for V2187 Cyg, that correspond to a B3 main sequence star. If the assumption is correct, then 
V2187~Cyg cannot be a subgiant, as many $\beta$ Cephei stars are. The fact that all photometric indices from 0.44 to 2.2 $\mu$m are well fitted 
by a single reddened star with the above properties impose  strong restrictions on a possible companion stellar object, as this would have to be 
much fainter  than the B3 star at all these wavelengths.

Finally, we derive the value of the ``pulsation constant'' $Q$ for V2187~Cyg,
$$Q = P \sqrt{\frac {\overline{\rho}} {\overline{\rho}_\odot}}$$ 
where $P$ is the period and $\frac{\overline{\rho}} {\overline{\rho}_\odot}$ is the mean star density in solar units. For a B3V
star, ${\rm log}~(\frac {\overline{\rho}} {\overline{\rho}_\odot}) = -1.15$ (Schmidt-Kaler, 1982; Section 4.1.5.2) and $P = 0.254$ days, 
then $Q = 0.07$ days, consistent with the values of $\beta$~Cephei variables in the catalog of Stankov \& Handler (2005). 
For a luminosity class IV, $Q$ would be slightly smaller.

\begin{table*}[htbp]
\small
  \setlength{\tabnotewidth}{0.9\textwidth} 
    \tablecols{8} 
    \caption{\label{tab3} Flux densities of V2187 Cyg at Other Wavelengths}
      \begin{center}
	    \begin{tabular}{cccccccc}\hline\hline 

  & JD & & Wavelength & Phase & Phase Calibrator & VLA & Source   \\
   Epoch & -2440000$^a$ &  Project & (cm) & Calibrator & Flux Density (Jy) & Conf.  & Flux Density (mJy)$^b$ \\
    \hline
     1984 Nov 28 & 06032.52   & AC116  & 20 &  2007+404 & 3.86$\pm$0.01 & A & $\leq$0.24$^c$ \\
     1993 May 01 & 09109.11 & AS483  & 3.6 & 2007+404 & 3.19$\pm$0.01 & B   & 1.05$\pm$0.11 \\
     2004 Feb 15 & 13051.28 & AS786  & 3.6 & 2015+371 & 2.72$\pm$0.01 & CnB & 1.15$\pm$0.13 \\
     2012 Jun 18$^d$ & 16096.92     & 12A-007   & 12  & 2007+404    & 0.930$\pm$0.002 &  B &  0.16$\pm$0.02 \\

   \hline\hline
  \tabnotetext{a}{At the middle of the observing run.}
 \tabnotetext{b}{Corrected for the primary beam response.}
 \tabnotetext{c}{Three-sigma upper limit.}
 \tabnotetext{d}{Observations made with the Jansky Very Large Array.}
   \label{tab3:1}
     \end{tabular}
     \end{center}
      \end{table*}

\begin{table*} [htbp]
\small
  \tablecols{5} 
\caption{\label{tab4} Mean FWHM of Balmer$^A$  and Helium$^B$ absorption lines.}
  \begin{center}
      \begin{tabular}{ccccc}\hline\hline   
      Phase $\phi$ &  \multicolumn {4} {c} {Width (\AA)}   \\
      (P=12.8 days) & \multicolumn {2} {c} {V2187 Cyg} &  \multicolumn {2} {c} {T91-507}  \\
       & H  &  HeI  & H & HeI \\
\hline
0.81   &   12.2 &  9.8   &     -   &   - \\     
0.17   &   12.6 &  9.7   &     -   &   - \\       
0.32    &  12.6 &  9.7   &     -   &   - \\
0.48   &   11.7 &  8.4   &     -   &   - \\
  -      &      -      &  -     &  8.9  &  5.1 \\
\hline\hline
  \tabnotetext{A}{H~(H$\alpha$,~H$\beta$~and~H$\gamma$)}
    \tabnotetext{B}{HeI~($\lambda\lambda$4438,4472,4931,5015,5875,6678)}
  \label{tab4:1}
   \end{tabular}
\end{center}
\end{table*}

\subsection{The Nature of the Radio Emission}

The radio emission from V2187 Cyg has a spectral index suggestive of 
partially thick free-free emission. One possible explanation could be 
that we are observing the emission from the stellar ionized wind. However, 
the expected flux densities at centimeter wavelengths of ionized winds from 
massive stars and the observed values  (0.4 to 0.8 mJy at 6 cm) are orders of magnitude 
higher than those expected for an early B-type star at a distance of 1.62 kpc  
(Dzib et al. 2013). 

Another possibility is that V2187 Cyg is ionizing a small HII region. 
The expected ionizing photon rate for a B2 V star is $7.8 \times 10^{44}~s^{-1}$ (Panagia 1973), 
that at a distance of 1.62 kpc will produce a flux density of $\sim$3 mJy at 3.6 cm. 
In other words, V2187 Cyg can provide the photoionization required to explain the 
radio continuum emission. However, the optical spectra discussed here does not 
present evidence of the emission lines expected from an HII region. 

The presence of a low-mass close companion whose neutral wind is, at some distance, ionized by 
UV photons  from V2187 Cyg may also  explain the observed radio emission.
The radio flux variations with a periodicity of 12.8 days would be caused by periodic eclipses of
this ionized region by the optically thick stellar wind of the primary star. Alternatively,
these may be caused by  periodic gas ejections by the secondary.

The main problem with the proposed mechanisms is that the optical observational results 
presented here do not support the presence of a binary system. In fact, no emission lines 
are present in the stellar spectrum, and no variations are seen in the profile of any absorption line. 
Furthermore, the photometry from 0.44 to 2.2 $\mu$m does not show any deviation from a single 
B2-3~V star reddened by $A_V = 5.8$. On the other hand, the object is also an X-ray source, which
would argue in favor of a binary scenario with a compact companion.

Finally, we consider the possibility that the radio emission is a fortuituous 
alignment with a background source. Following Windhorst et al. (1993),  the 
\sl a priori \rm probability of finding a 3.6 cm source of 0.4 mJy in a solid angle 
of 1 square arcsecond is only $\sim 2 \times 10^{-6}$, so we consider this possibility 
very unlikely. 

\subsection{A search for radio continuum from other $\beta$ Cephei stars}

The catalog of galactic $\beta$ Cephei stars of Stankov \& Handler (2005) lists a total of 93 stars.
More recently, Pigulski \& Pojma{\'n}ski (2008) reported an additional 103 $\beta$ Cephei stars.
We searched for VLA archive observations of good quality toward these stars.

We found observations for only 14 stars in the Stankov \& Handler (2005) catalog and only for two stars in the Pigulski
\& Pojma{\'n}ski (2008) catalog.  The small number of stars with VLA observations in this last, more recent, catalog 
is because most of the stars in it are very southern objects. Many of these observations were pointed to 
other objects, but included the $\beta$ Cephei star in the primary beam. 

We set as a criterion that the final image was expected to provide an rms noise in the range of 0.1 mJy and thus a 
detection at the mJy level.  Of these 16 $\beta$ Cephei stars, only  V2187 Cyg was detected as a radio source.
The upper limits for the other stars are listed in Table 5.  It must be stressed that not even
the prototype of the class, the third-magnitude star $\beta$ Cep, with a Hipparcos measured paralax of 4.76 mas 
($d = 210$ pc) could be detected at 3.6 cm, with a 3-$\sigma$ upper limit  of 0.15 mJy.

\begin{table*}[htbp]
\small
  \setlength{\tabnotewidth}{0.9\textwidth} 
    \tablecols{9} 

  \caption{\label{tab} Upper limits to $\beta$ Cephei Stars observed with the VLA}
  \begin{center}
    \begin{tabular}{lcccccccc}\hline\hline
& Other & \multicolumn{2}{c}{Position{$^a$}} & & & VLA & $\lambda$ & Upper  \\
\cline{3-4} 
Star & Name &  $\alpha$(J2000) & $\delta$(J2000) & Epoch & Project & Conf. & (cm) & Limit (mJy){$^b$} \\ 
\hline
$\gamma$ Peg & HD 886 & 00 13 14.15 & +15 11 00.9 & 1992 Apr 14 & AD288 & C & 3.6 & $\leq$0.20 \\
V619 Per & SAO 23246 & 02 22 02.78 & +57 08 25.1 & 1984 Sep 29 & AR110 & D & 6 & $\leq$0.17 \\
V595 Per & SAO 23251 & 02 22 08.61 & +57 07 28.4 & 1984 Sep 29 & AR110 & D & 6 & $\leq$0.17 \\
V1032 Sco & Braes 930 & 16 53 58.60 & --41 48 41.8 & 1990 Feb 17 & AH395 & A & 3.6 & $\leq$0.17 \\
V946 Sco & Braes 932 & 16 54 01.77 & --41 51 12.1 & 1990 Feb 17 & AH395 & A & 3.6 & $\leq$0.14 \\
V964 Sco & Braes 672 & 16 54 18.32 & --41 51 35.7 & 1990 Feb 17 & AH395 & A & 3.6 & $\leq$0.17 \\
HD 167451 & SAO 161234 & 18 15 54.86 & --13 34 27.7 & 1997 Oct 11 & AK451 & DnC & 21 & $\leq$0.60 \\ 
HD 172427 & SAO 161713 & 18 40 32.18 & --10 43 06.9 & 1995 Jan 21 & AR330 & DnC & 6 & $\leq$0.17 \\
NSV 13054 & NGC 6910 18 & 20 22 58.94 & +40 45 39.3 & 1994 Oct 11 & AS544 & C & 6 & $\leq$0.42 \\
NGC 6910 16 & --- & 20 23 07.34 & +40 46 55.6 & 1994 Oct 11 & AS544 & C & 6 & $\leq$0.33 \\
TYC 3156-1857-1 & NGC 6910 14 & 20 23 07.57 & +40 46 08.9 & 1994 Oct 11 & AS544 & C & 6 & $\leq$0.33 \\
TYC 3156-1028-1 & NGC 6910 27 & 20 23 33.75 & +40 45 19.9 & 1994 Oct 11 & AS544 & C & 6 & $\leq$0.39 \\ 
BW Vul & HD 199140 & 20 54 22.39 & +28 31 19.2 & 1986 Jun 19 & AH204 & A & 6 & $\leq$0.07 \\
$\beta$ Cep & HD 205021 & 21 28 39.60 & +70 33 38.6 & 2002 Jun 02 & AD467 & A & 3.6 & $\leq$0.15 \\
DD Lac & HD 214993 & 22 41 28.65 & +40 13 31.6 & 1992 Apr 14 & AD288 & C & 3.6 & $\leq$0.20 \\

\hline\hline
\tabnotetext{a}{From SIMBAD. The position of V595 Cyg in the catalog of Stankov \& Handler (2005) differs
by $\sim 1'$ from that given in SIMBAD.} 

\tabnotetext{b}{At the 3-$\sigma$ level.}
    \label{tab:2}
    \end{tabular}
  \end{center}
\end{table*}

\section{Conclusions}

1. We report the detection of centimeter radio continuum emission toward 
the $\beta$ Cephei star V2187 Cyg. The 6 cm emission is faint (0.4 to 0.8 mJy) 
and shows a periodicity of around 12.8 days. We searched unsuccessfully for radio 
emission toward other 15 $\beta$ Cephei stars with VLA archive observations. 

2. We present extensive new optical observations of V2187 Cyg that 
support its classification as a $\beta$ Cephei star based on sinusoidal 
light variations with a confirmed period of 0.25389 days and mean amplitudes 
of 0.37 and  0.42 magnitudes in $i$ and $V$ bands, respectively. 

3. We classify the spectrum of V2187 Cyg as B3~V with no evidence of a stellar
companion or any emission lines in the spectral range 4300 to 6800 \AA.

\acknowledgments

LFR acknowledges the support of DGAPA-UNAM, and of CONACyT (M\'exico).
MT acknowledges the support  from grant PAPIIT/DGAPA IN-101813.
GT, SZ and DGB are supported  by CONACyT grant 166376 and PAPIIT/DGAPA projects IN-107712 and IN-103912.
This research has made use of the SIMBAD database, operated at CDS, Strasbourg, France.

\end{document}